\documentclass[review]{elsarticle}

\usepackage{hyperref}

\journal{Advances in Space Research}

\biboptions{authoryear}

\begin{document}

\begin{frontmatter}

\title{Spectral fluctuation analysis of ionospheric inhomogeneities over Brazilian territory \\Part II: E-F valley region plasma instabilities}

\author{{\normalsize Neelakshi J.}$^{1*}$}
\cortext[cor]{Corresponding author: neelakshij@gmail.com }
\author{Reinaldo R. Rosa$^{1}$}
\author{Siomel Savio$^{2,3,4}$}
\author{Francisco C. de Meneses$^{3,5}$}
\author{Stephan Stephany$^{1}$}
\author{Gabriel Fornari$^{1}$}
\author{Muralikrishna P.$^{2}$}

\address{$^{1}$Computational Space Physics Group, Lab for Computing and Applied Math (LABAC), National Institute for Space Research (INPE), Av. dos Astronautas, 1758, S\~ao Jos\'e dos Campos, S\~ao Paulo, 12227-690, Brazil}
\address{$^{2}$Aeronomy Division, National Institute for Space Research (INPE), Av. dos Astronautas, 1758, S\~ao Jos\'e dos Campos, S\~ao Paulo, 12227-690, Brazil}
\address{$^{3}$China-Brazil Joint Laboratory for Space Weather, NSSC/INPE, Av. dos Astronautas, 1758, S\~ao Jos\'e dos Campos, S\~ao Paulo, 12227-690, Brazil}
\address{$^{4}$State Key Laboratory of Space Weather, National Space Science Center (NSSC), Chinese Academy of Sciences, No. 1, Nanertiao, Zhongguancun, Haidian District, Beijing 100190, China}
\address{$^{5}$School of Physics and Mathematics, Autonomous University of Nuevo Le\'on (UANL), Av. Universidad s/n, Cd. Universitaria, San Nicol\'as de los Garza, N.L. 66455, Mexico}

\begin{abstract}

The turbulent-like process associated with ionospheric instabilities, has been identified as the main nonlinear process that drives the irregularities observed in the different ionospheric regions. In this complementary study, as proposed in the first article of this two-paper series [Fornari et al., Adv. Space Res. 58, 2016], we performed the detrended fluctuation analysis of the equatorial E-F valley region electron density fluctuations measured from an in situ experiment performed over the Brazilian territory.
The spectral consistency with the K41 turbulent universality class is analyzed for E-F valley region from the DFA spectra for four electron density time series. A complementary detrended fluctuation analysis for four time series of the F-layer electric field is also presented. Consistent with the results obtained for the F region, the analysis for the E-F valley region also shows a very high spectral variation ($\gg50\%$). Thus, the spectral analysis performed in both parts of the series suggest that a process such as the homogeneous turbulence K41 ($\beta =-5/3\pm 2\%$) is inappropriate to describe both the fluctuations of electron density and the electric field associated with the main ionospheric instabilities.

\end{abstract}


\end{frontmatter}


\section{Introduction}

The characteristic features of in situ ionospheric plasma density fluctuation data may provide important information on the structural processes associated with ionospheric irregularities \citep{Muralikrishna2003}. In Part I of this work, \citet{Fornari2016} analyzed in situ F region electric field fluctuation data using the detrended fluctuation analysis (DFA) \citep{Peng1994} technique to verify the wide variation in the spectral indices reported in earlier rocket experiments based on power spectral density (PSD) method. The results show that the high variability of the spectral indices is not due to the statistical limitation of the data, and does not constitute a K41 type of universality class \footnote{The K41 is a theoretical framework for turbulence proposed by Kolmogorov in 1941, which forms a basis to understand the behavior of homogeneous multiplicative energy cascade from turbulent-like processes. Here the turbulent energy spectrum follows a precise power law behavior with index $-5/3\pm 2\%$ in the inertial range \citep{Frisch1995}. Therefore, the K41 spectrum represents a universality class for homogeneous turbulent processes.}. 
As shown, in the first part of this study, PSD although widely used, falls short to characterize turbulence spectra from in situ ionospheric plasma density fluctuation measurements. Many studies have shown that the power spectra of these fluctuations exhibit two or three different spectral exponents indicating the scaling complexity of the process involved \citep{Kelley1991, Spicher2014}. In general, the spectral indices that have been reported show deviation from the K41 theory but do not elucidate the statistical properties of the energy cascading that is supposed to drive the  ionospheric turbulence \citep{Kelley1991}. In this context, the DFA proposed by \citet{Peng1994} is a potential method that could render insights into the statistical properties of the turbulence phenomena.

As envisioned in Part I, here (Part II) the DFA is applied to in situ E-F valley region (hereafter, valley region) data. The valley region is located between the top of the E region and the base of the F region. The valley region, specifically the equatorial ones, hosts a variety of plasma irregularities both during the day, the so-called 150 km echoes \citep{Kudeki1993,Rodrigues2011}, and at dusk-nighttime \citep{Chau2004}. This region is still a less explored area of research compared to the F region given the technical limitations in observing it. It can be studied by using powerful incoherent and coherent scatter radar and in situ experiments. Various studies have been reported on the correlation between the valley region irregularities and the equatorial plasma instabilities in the F region:

 \begin{itemize}
     \item Radar observations revealed that (i) the valley region irregularities are often found when the equatorial spread F (ESF) occurred after the sunset and that their spatial structures and temporal variations have resemblance with the ESF, and (ii) the valley region irregularities are a result of the coupling between the unstable equatorial F region and the underlying low-latitude valley and the E region \citep{Vickrey1982,Vickrey1984,Patra2008,Yokoyama2005,Li2011,Kherani2012}.
     \item Studies based on in situ data found that electric field and gravity waves may play a key role in the generation of these structures (in the valley regions) and that the structures are produced by the generalized Rayleigh-Taylor instability mechanism at the base of the F region \citep{Vickrey1984, Prakash1999, Sinha1999, Muralikrishna2003, Savio2017}. \citet{Savio2017} reported the presence of wave-like structures in valley region data obtained from an experiment over Brazil, and the same data is used for the present analysis.
 \end{itemize}
 
In literature, the DFA is applied to study ionospheric irregularities, but we could not find its application to in situ valley region data. This work presents the first instance of application of the DFA to in situ E-F valley region electron density fluctuation data. The paper is organized as follows. Section 2 describes the data along with the electron density vertical profile. The DFA is presented in Section 3 followed by the concluding remarks in Section 4.

\section{In situ valley region data}   

\begin{figure}
\centering
\includegraphics[width=1.0\linewidth, height=1.0\linewidth] {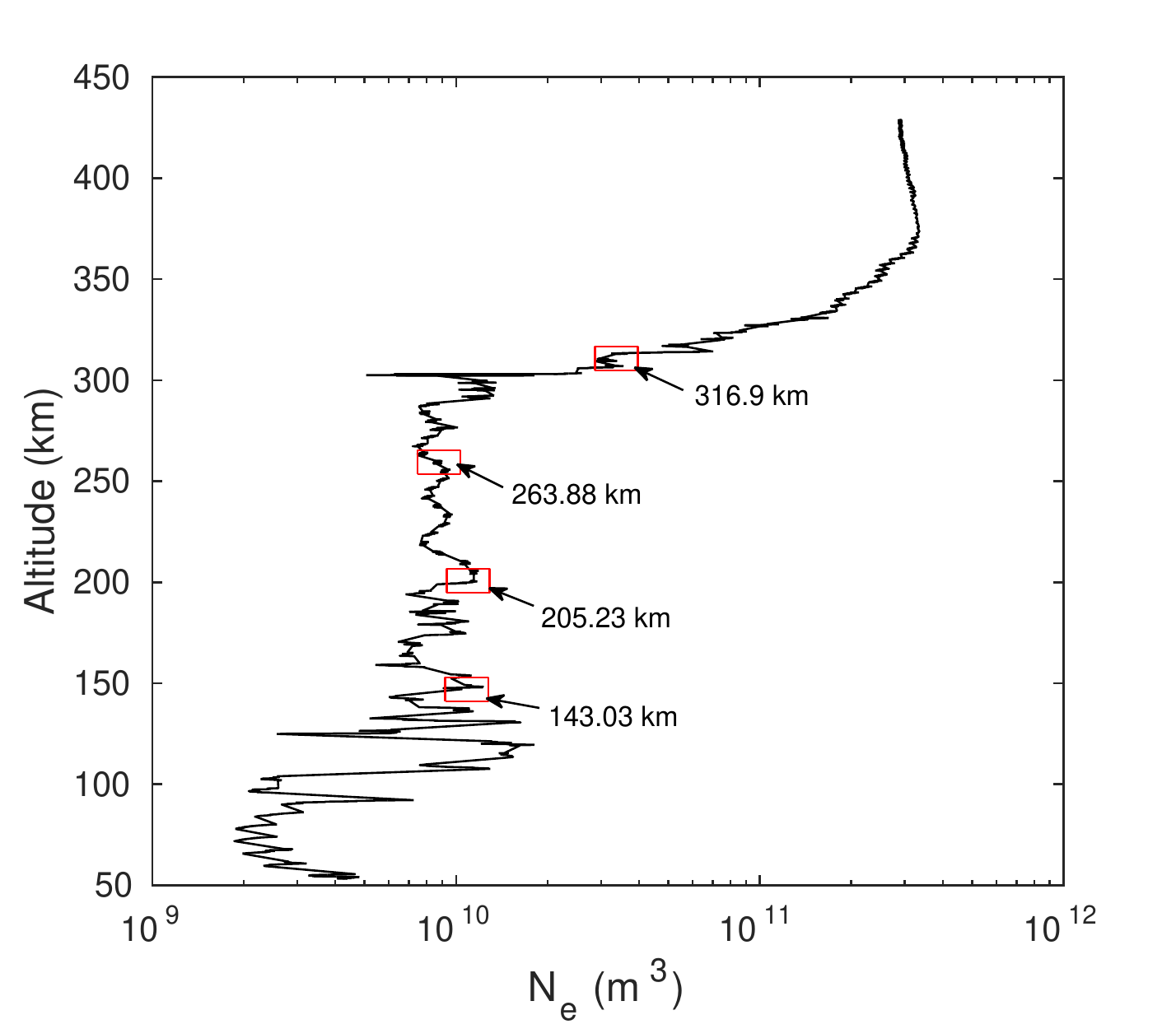}
\caption{Vertical profile of electron densities for downleg trajectory. Open red boxes represent the chosen heights.}
\label{fig:profile}
\end{figure}

The vertical profile of electron density was obtained from a conical Langmuir probe on-board a two-stage VS-30 Orion sounding rocket experiment launched from an equatorial rocket launching station, Alc\^antara ($2.24^{\circ}$ S, $44.4^{\circ}$ W, dip latitude $5.5^{\circ}$ S), on December 8, 2012, at 19:00 LT, under quiet geomagnetic conditions. During the $\sim$11 min flight, rocket trajectory was in the north-northeast direction towards the magnetic equator, ranging $\sim$384 km horizontally with an apogee covering typical F region altitudes of $\sim$428 km.
The conical Langmuir probe worked both in swept and constant bias modes. The probe sensor potential was swept from -1 V to +2.5 V linearly in about 1.5 s, during which the electron kinetic temperature was determined from the collected probe current. Then, the potential was maintained at +2.5 V (constant bias mode) for 1 s, during which the collected probe current was used to estimate electron density and its fluctuations, in each experiment cycle. This work utilizes the electron density fluctuation data obtained from the conical Langmuir probe. Fig. \ref{fig:profile} shows variations in the vertically distributed electron density in the downleg (descent of the rocket) trajectory of the flight. 

At the time of launch, the ground-based equipment detected conditions favorable for the generation of plasma bubbles in the F region.  \citet{Savio2017} reported the presence of several small- and medium-scale plasma irregularities in the valley region (120-300 km) during both ascent and descent, which were more prominent during the descent of the rocket. In the downleg profile, the average electron density observed was around $9\times10^9$ $m^{-3}$, equivalent to $1/10$th of the E region maximum, and then, it gradually increased after $300$ km, where the broad base of F region was detected. These observations are consistent with the work reported by \citet{Wakai1967}, which stated that under quiet conditions, the electron concentration in the valley around midnight is about $1/10$th of the E region maximum, and width of the valley is very wide compared to the disturbed nights. \citet{Prakash1970} reported observing a deep valley region above $120$ km, i.e., $120$-$140$ km, where the electron density fell by two orders of magnitude in their experiment (to a few hundreds per cubic centimeters).
Fig. \ref{fig:valleyRegion} (left panel) shows the selected time series from the downleg electron density profile around average heights of 143, 205, and 263.9 km from the valley region and around 316.9 km, just above the wide base of the F region.

\section{Fluctuation analysis, results and interpretation}

The DFA proposed by \citet{Peng1994} could render insights into the statistical properties of turbulence phenomena. Originally proposed to detect long-range correlations in DNA sequences and in data influenced by trends, the DFA is widely used in many branches of sciences - medicine, physics, finance and social sciences - to understand the complexity of systems through its scaling exponent that characterizes fractal dynamics of the system \citep{Kantelhardt2009, Veronese2011}.

The robustness of DFA can be attributed to some of its interesting features.  For instance, \citet{Coronado2005} investigated the influence of the length of a time series in quantifying the correlation behavior using techniques like autocorrelation analysis, Hurst exponent, and DFA. The comparison study revealed that the DFA is practically unaffected by the length of time series, contrary to that observed from the results of Hurst analysis or autocorrelation analysis. Another interesting feature has been reported by \citet{Chen2002} who altered time series by excluding parts of it, stitching the rest and subjecting it to the DFA. The study revealed that even with the removal of 50\% of the time series, the scaling behavior of positively correlated signals is unaltered, implying that time series need not be continuous. \citet{Heneghan2000} established an equivalence relation between the PSD exponent, $\beta$, and the DFA exponent, $\alpha$, given by $\beta \equiv 2\alpha-1$. \citet{Kiyono2015} showed that this relationship is valid for the higher order DFA subject to the constraint $0<\alpha<m+1$, where $m$ is the order of detrending polynomial in the DFA.

\begin{figure}
\centering
\includegraphics[width=1.0\linewidth, height=1.4\linewidth] {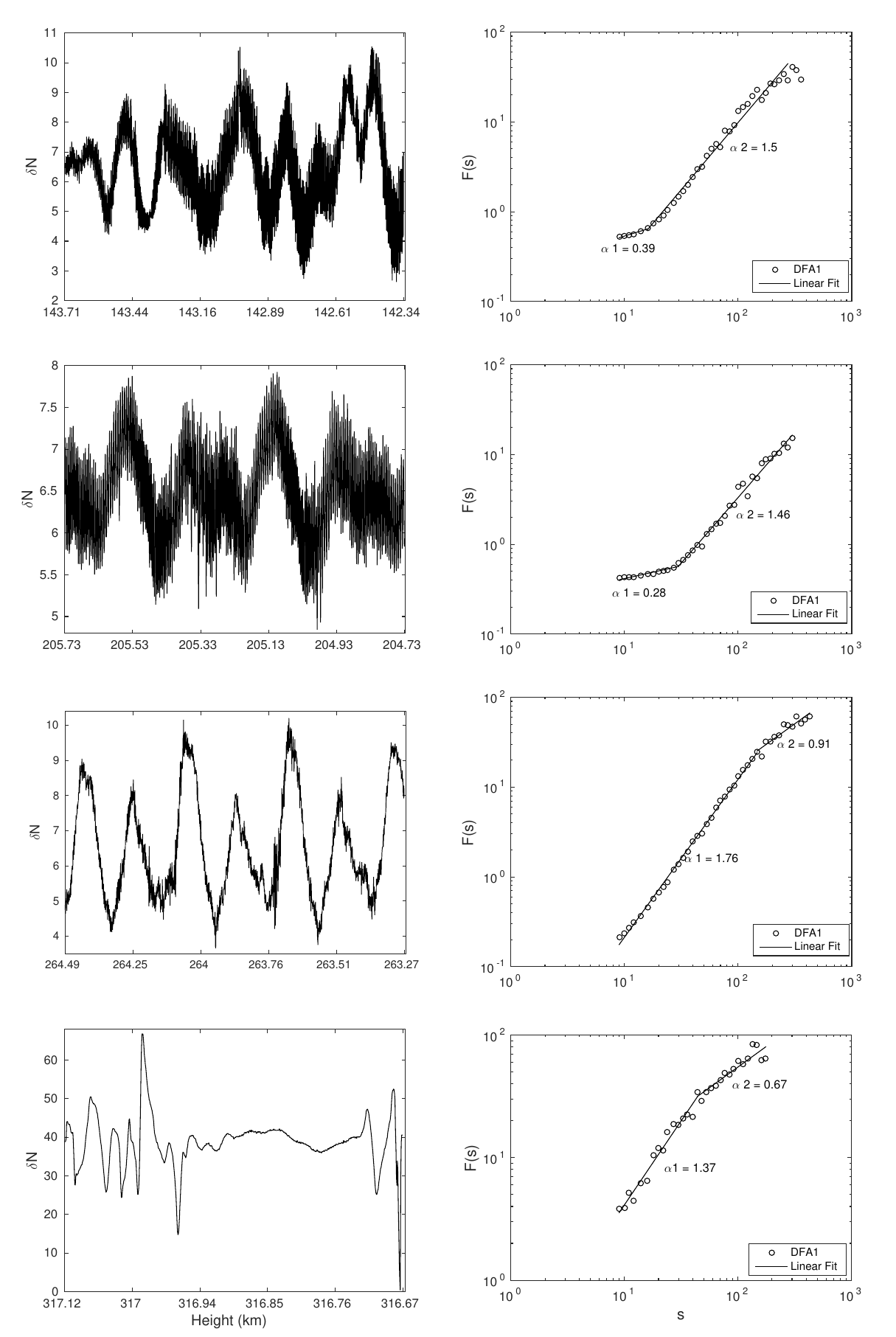}
\caption{Left column - E-F valley region time series ($\delta N$) of electron density fluctuations obtained from Langmuir probe during the downleg flight for the chosen heights. Right column - Corresponding fluctuation function profile $F(s)$ (open circle) as a function of scales $s$ along with the fit (solid line), i.e., the $\alpha$ exponent.}
\label{fig:valleyRegion}
\end{figure}

The DFA involves obtaining cumulative sum of the mean subtracted time series followed by dividing it into non-overlapping segments $(s)$, referred to as scales. Further, these segments are detrended using the linear least squares or higher order polynomial ($m$) method and the variance is calculated. Depending on the detrending order, $m$, the analysis is referred to as DFA$m$. Averaging the root mean square over the segments $(s)$ gives the fluctuation function, $F(s)$. Linear fit to the fluctuation function profile yields the scaling exponent $\alpha$. Implementation procedure can be found in Part I of this paper \citep{Fornari2016}. In this work, four time series of electron density fluctuations from the downleg profiles corresponding to the valley region are selected. The selected time series correspond to the mean heights of $143, 205, 263.9,$ and $316.9$ km (please see left panel in Fig. \ref{fig:valleyRegion}).

The selected time series are subjected to DFA. Scales are varied from $10$ to $N/4$ with a factor of $2^{\frac{1}{8}}$, where $N$ is the length of time series \citep{Goldberger2000}. The fluctuation function computed from DFA is plotted as a function of scales for all the selected time series (right panel in Fig. ~\ref{fig:valleyRegion}) on a log-log scale. The profiles of fluctuation function for all the chosen cases exhibit long-range correlation with a crossover. Crossover refers to a change in the scaling exponent for different scale ranges, and it usually arises due to a change in the correlation properties over different spatial or temporal scales, or from trends in the data. The exponents $\alpha1$ and $\alpha2$ are obtained from the linear fit of $F(s)$, where $\alpha1$ refers to smaller scales and $\alpha2$ refers to larger scales. Our analysis reveals $\alpha1$ to be in the range of $0.28$ to $1.76$ and $\alpha2$ in the range of $0.67$ to $1.5$. For mean heights corresponding to $143$ and $205$ km, we observe $\alpha1$ is smaller than $\alpha2$, contrary to the observation for mean heights corresponding to $263.9$ and $316.9$ km. 

\begin{figure}
    \centering
    \includegraphics[width=1.0\linewidth, height=1.0\linewidth] {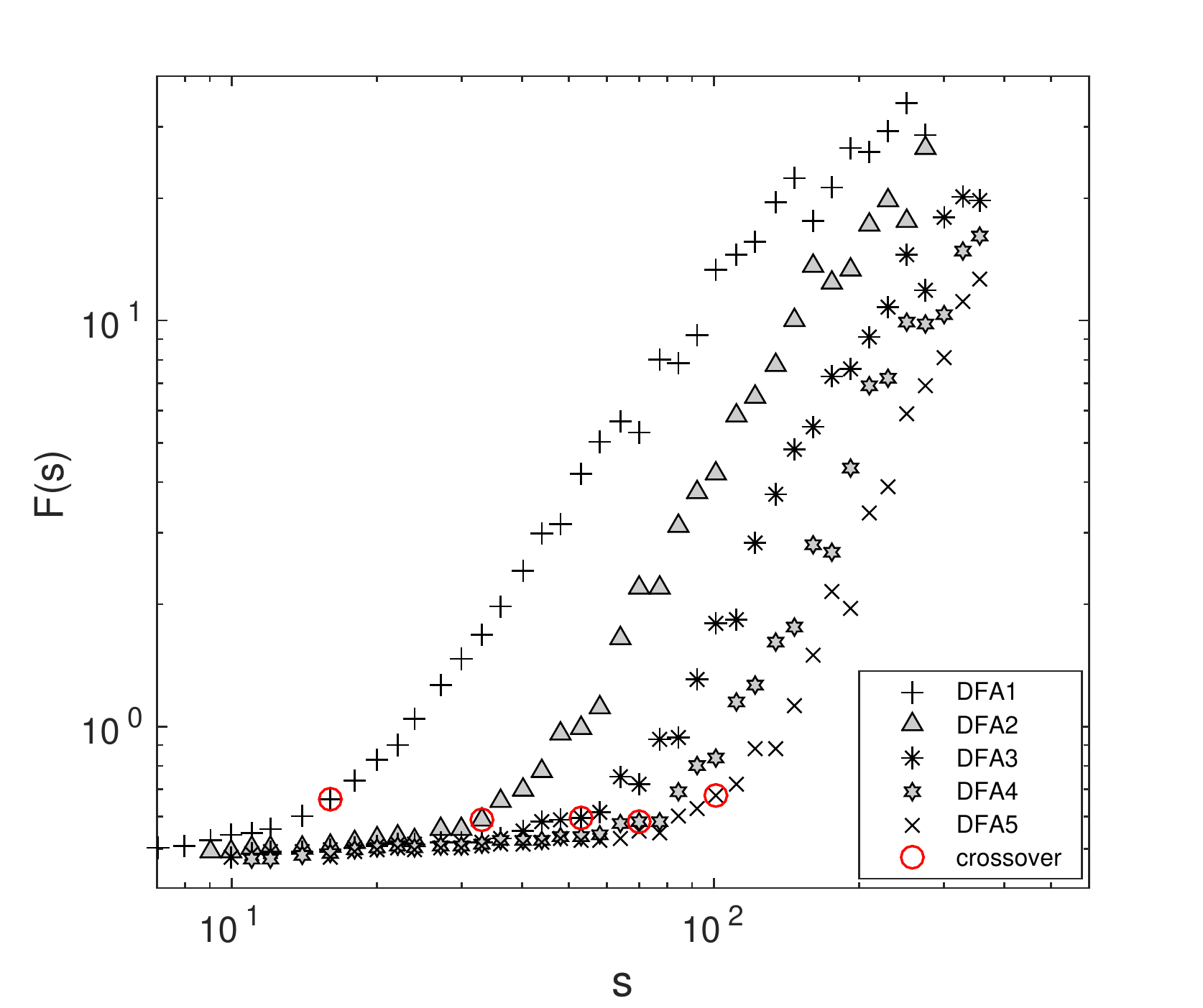}
    \caption{Fluctuation function profiles for the downleg time series at a mean height of 143 km for polynomials of orders $1$-$5$. Open red circles represent the crossover points for the respective detrending order.}
    \label{fig:crossover}
\end{figure}

In order to be sure that the obtained crossover is intrinsic to the data and not an artifact, we investigated the time series with higher order DFAs, of the order $1$-$5$. For this investigation, we used the methodology prescribed by \citet{Kantelhardt2001} to identify false crossovers. Artificial crossover exhibits similar characteristic length with identical scaling. Fig. ~\ref{fig:crossover} presents the analysis for downleg time series corresponding to the mean height of $143$ km with DFA of $1^{st}$ to $5^{th}$ order. The crossover exponents are listed in Table \ref{tab1:crossover}. It can be observed that as the order of detrending increases, crossover point moves towards larger scales and have different scaling exponents. This investigation confirms that the obtained crossover is an intrinsic property of electron density fluctuation data in the valley region.

The PSD exponent, $\beta$, is calculated using the equivalence relationship given above, and the standard deviation $\sigma_m$ (in $\%$) is determined. The computed DFA exponents in our analysis show a wide range of $\beta$ from $-0.98$ to $-2.14$ with $\sigma_m = 58\%$. Table \ref{tab2:deviation} summarizes the variations in the $\beta$ exponent obtained from the previous equivalent studies \citep{Rino1981, Kelley1982, Muralikrishna2007, Sinha2010, Sinha2011} and compares with the present work. All studies reported in Table \ref{tab2:deviation} are based on electron density fluctuation data obtained through rocket experiments. It is observed that the computed standard deviation $\sigma_m\gg50\%$, which affirms that the underlying mechanism for instabilities differs from the K41 homogeneous turbulence, given the accepted deviation is $\sigma_m\le2\%$ \citep{Frisch1995}.

\begin{figure}
\centering
\includegraphics[width=1.0\linewidth, height=1.0\linewidth] {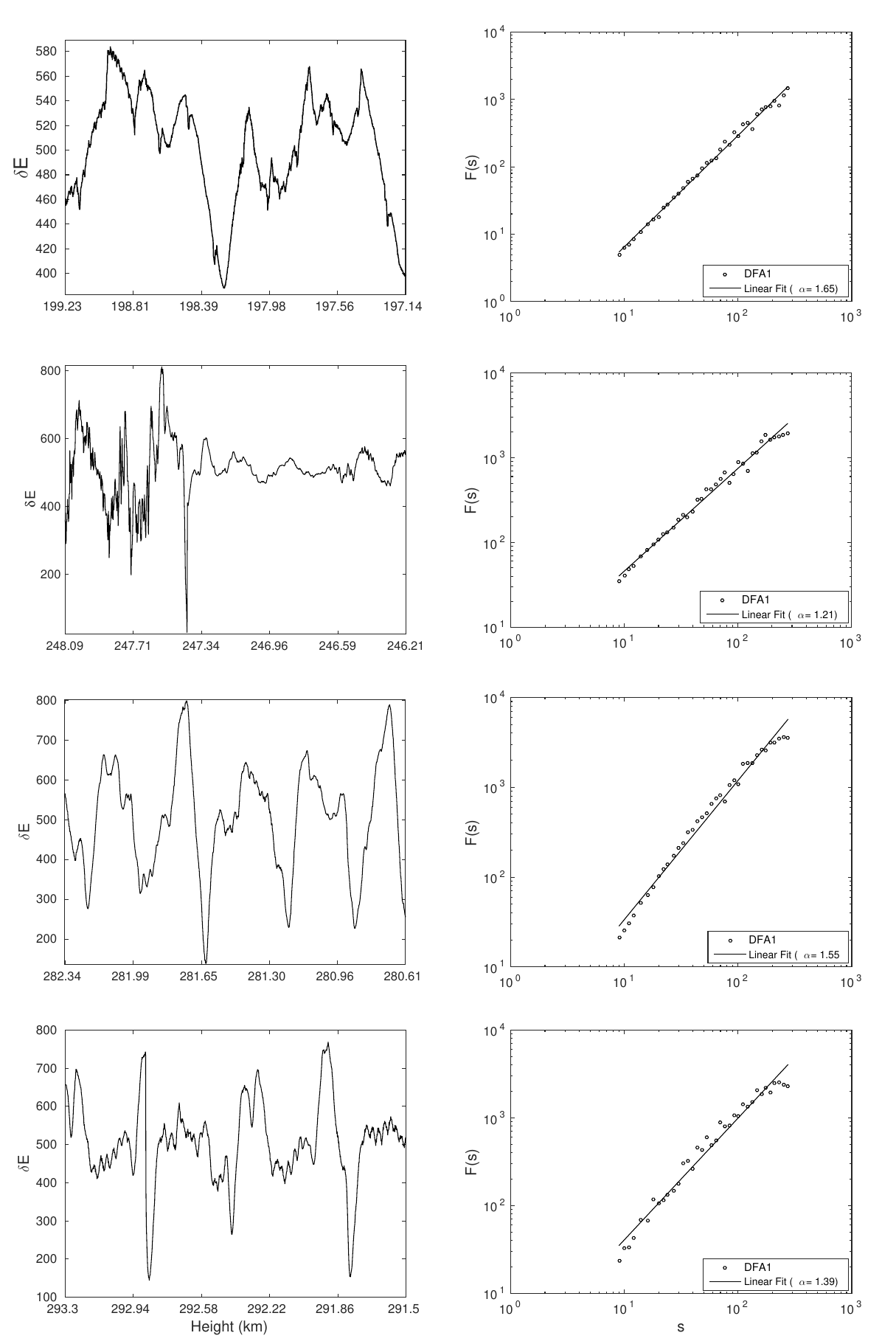}       
\caption{Left column - F region time series ($\delta E$) of electric field fluctuations obtained from electric field probe during the downleg flight. Right column - Corresponding fluctuation function profile $F(s)$ (open circle) as a function of scales $s$ along with the fit (solid line), i.e., the $\alpha$ exponent.}
\label{fig:fregion}
\end{figure}

We also performed the DFA on in situ electric field fluctuation data from the F region obtained from an earlier experiment conducted on December 18, 1995, at 21:17 LT, under quiet geomagnetic conditions from the same equatorial launching station Alc\^antara (2.24$^{\circ}$ S, 44.4$^{\circ}$ W, dip latitude 5.5$^{\circ}$ S) \citep{Fornari2016}. The rocket flight traversed through similar altitudes of $200$-$300$ km. This data indicated the presence of a large plasma bubble at an altitude of $\sim$280 km. Fig. \ref{fig:fregion} presents the time series and the corresponding DFA.

\begin{figure}
\centering
\includegraphics[width=1.0\linewidth]{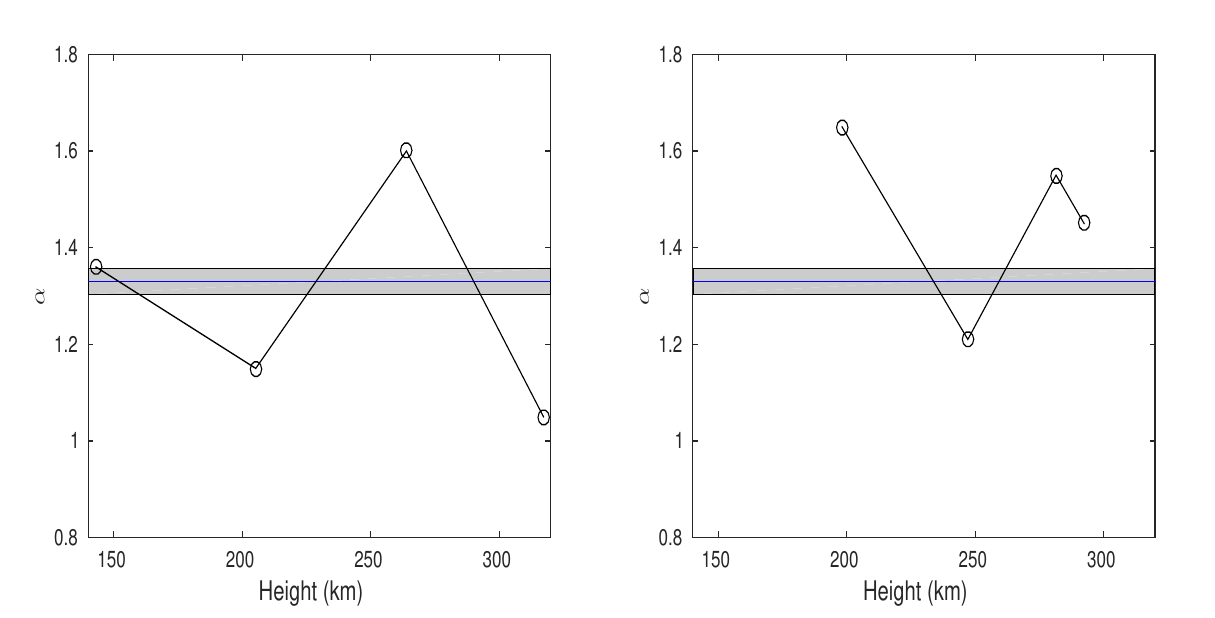}    
\caption{Mean height vs. DFA exponent $\alpha$ for downleg plasma density fluctuation time series: E-F valley region (left) and F region (right). Solid line in the shaded area indicates the exponent value for homogeneous turbulence (k41 theory) with $\beta=1.66$, i.e., $\alpha=1.33$; shaded area shows the range of alpha value deviation $\pm2\%$.}
\label{fig:alpha_height}
\end{figure}

The data from aforementioned experiments is selected for the altitudes of $200$-$300$ km. In valley region data, small-to-medium scale plasma irregularities \citep{Savio2017} are found, while F-region data shows medium-to-large scale plasma irregularities (Fig. 2 in \citeauthor{Muralikrishna2003}, 2003). Hence, it will be interesting to compare the scaling exponents of plasma densities around similar altitudes for these two different regions. Fig. \ref{fig:alpha_height} shows the scaling exponent plotted as a function of height for the  valley region (left panel) and the F region (right panel). For this plot, we have used a single linear fit for valley region data. The shaded horizontal bar in the plot represents the exponent value, $\alpha =1.33\pm 2\%$, for the homogeneous turbulence described by the K41 theory. The range of $\alpha$ exponents for the F region is higher than that of the valley region, which may be due to different scaling present in these regions. Wide variations of the scaling exponent from the K41 theory are observed for both regions.

\section{Concluding remarks}

In this paper, the complementary in situ E-F valley region irregularities are studied using the DFA. This study is important as studies of the equatorial E-F valley region at nighttime are scarce. Our analysis shows that the E-F valley region electron density fluctuations exhibit long-range correlation with crossovers that are intrinsic to the data for all the chosen altitudes. The F region irregularities obtained from an earlier experiment are also analyzed using the DFA and similar results in terms of long-range correlations are obtained for all the chosen altitudes. The PSD exponent $\beta$ is computed for the current data and compared with earlier similar experiments. The results show $\sigma_m\gg50\%$. These observations along with the profile of $\alpha$ with respect to the height indicate that scaling exponents show wide variation from the K41 theory, for both the E-F valley and F regions. This implies that the turbulent-like ionospheric fluctuations as a whole cannot be described by the K41 homogeneous energy cascade theory.

Given this scenario and considering the different mechanisms responsible for the plasma instability along different ionospheric regions, it is necessary to investigate the model for non-homogeneous turbulence that will help to understand the observed high spectral variability. A future study that emerges naturally in this scenario is to look for multifractal signature from the data analyzed here. This investigation is in progress and will be published in an upcoming paper.

\section*{Acknowledgement}
The authors are grateful to the Institute of Aeronautics and Space (IAE/DCTA) and Alc\^{a}ntara Launch Center (CLA) for providing sounding rocket and launch operation, respectively.
NJ acknowledges the financial assistance received from CAPES. 
RRR is grateful to FAPESP sponsored by Process No. $2014/11156-4$.
S.Savio acknowledges the financial support from China-Brazil Joint Laboratory for Space Weather, National Space Science Center, Chinese Academy of Science. 
F.C. de Meneses acknowledges the financial support given by the National Council of Science and Technology of Mexico (CONACYT), CAS-TWAS Fellowship for Postdoctoral and Visiting Scholars from Developing Countries under Grant no. 201377GB0001, and the Brazilian Council for Scientific and Technological Development (CNPq) under Grant number $312704/2015-1$. 
S.Stephany thanks CNPq for grant 307460/2015-0

\clearpage  


\begin{table}
\caption{DFA1 to DFA5 for downleg time series at $\sim$143.03 km.}
\label{tab1:crossover}
\centering
\begin{center}
    \tabcolsep=0.5cm
    {
    \begin{tabular}{|c|c|c|}
    \hline
        DFA order & $\alpha1$ & $\alpha2$ \\[0.5em] \hline
        DFA1 & 0.39 & 1.50 \\ [0.2em]\hline
        DFA2 & 0.15 & 1.93 \\ [0.2em]\hline
        DFA3 & 0.12 & 2.08 \\ [0.2em] \hline
        DFA4 & 0.10 & 2.26 \\ [0.2em] \hline
        DFA5 & 0.13 & 2.53 \\ 
    \hline    
    \end{tabular}
    }
\end{center}    
\end{table}

\begin{table}
\caption{Comparison of PSD spectral indices ($\beta$) found in previous equivalent studies and $\beta$ obtained here from DFA. All results measured using rockets are related to electronic density measurements during the experiment.}
\label{tab2:deviation}
\begin{center}
\tabcolsep=0.1cm
\scalebox{0.72}
{
\begin{tabular}{|c|c|c|c|c|c|c|c|c|}
\hline
Date and Time & Spacecraft & Altitude (km) & $\beta$ range & $\left\langle \beta\right\rangle$  & $\sigma_{m}$ & References \\ \hline
17/07/1979, 12:31:30 UT & Rocket & 250 to 370 & -1.20 to -3.4 & -2.3 & 110\%  & \cite{Rino1981}\\ \hline
17/07/1979, 12:31:30 UT & Rocket & 250 to 285 & -2.00 to -3.4 & -2.7 & 70\%  & \cite{Kelley1982}\\ \hline
11/12/1985, 00:30 UT & Rocket & 210 to 306 & -1.34 to -3.3 & -2.32 & 98\% & \cite{Muralikrishna2007}\\ \hline
31/10/1986, 03:00 UT & Rocket & 100 to 220 & -1.54 to -3.30 & -2.42 & 88\% & \cite{Muralikrishna2007}\\ \hline
14/10/1994, 22:55 UT & Rocket & 117 to 518 & -1.20 to -5.3 & -3.25 & 205\% & \cite{Muralikrishna2007}\\ \hline
18/12/1995, 00:17 UT & Rocket & 240 to 500 & -1.11 to -4.90 & -3.01 & 189\% & \cite{Muralikrishna2007}\\ \hline
15/01/2007, 16:43 UT & Rocket & - to 127 & -1.60 to -2.70 & -2.15 & 55\% & \cite{Sinha2010}\\ \hline
29/01/2008, 15:49 UT & Rocket & - to 117 & -2.00 to -3.50 & -2.75 & 75\% & \cite{Sinha2011}\\ \hline
08/12/2012, 22:00 UT & Rocket & - to 317 & -0.98 to -2.14 & -1.56 & 58\% & This paper\\ \hline
\end{tabular}
}
\end{center}
\end{table}

\clearpage

\end{document}